# Graph Partitioning using Quantum Annealing on the D-Wave System

**Hayato Ushijima-Mwesigwa†‡, Christian F. A. Negre§, and Susan M. Mniszewski‡§**

† School of Computing, Clemson University, Clemson, SC 29634, USA

‡ Computer, Computational, & Statistical Sciences Division, Los Alamos National Laboratory, Los Alamos, NM 87545, USA

§ Theoretical Division, Los Alamos National Laboratory, Los Alamos, NM 87545, USA

**Abstract.** In this work, we explore graph partitioning (GP) using quantum annealing on the D-Wave 2X machine. Motivated by a recently proposed graph-based electronic structure theory applied to quantum molecular dynamics (QMD) simulations, graph partitioning is used for reducing the calculation of the density matrix into smaller subsystems rendering the calculation more computationally efficient. Unconstrained graph partitioning as community clustering based on the modularity metric can be naturally mapped into the Hamiltonian of the quantum annealer. On the other hand, when constraints are imposed for partitioning into equal parts and minimizing the number of cut edges between parts, a quadratic unconstrained binary optimization (QUBO) reformulation is required. This reformulation may employ the graph complement to fit the problem in the *Chimera graph* of the quantum annealer. Partitioning into 2 parts, $2^N$ parts recursively, and $k$ parts concurrently are demonstrated with benchmark graphs, random graphs, and small material system density matrix based graphs. Results for graph partitioning using quantum and hybrid classical-quantum approaches are shown to equal or out-perform current "state of the art" methods.

§ To whom correspondence should be addressed (smm@lanl.gov)

## 1. Introduction

Quantum annealing (QA) is a combinatorial optimization technique meant to exploit quantum-mechanical effects such as tunneling and entanglement [1] to minimize and sample from energy-based models. Machines performing QA at the hardware level such as the D-Wave computer have recently become available [2, 3], and evidence for quantum mechanical effects playing a useful role in the processing have been seen [4, 5, 6]. The D-Wave system minimizes the following Ising objective function.

$$O(\mathbf{h}, \mathbf{J}, \mathbf{s}) = \sum_i h_i s_i + \sum_{i<j} J_{ij} s_i s_j \tag{1}$$

This is closely related to the Ising model energy function as a problem Hamiltonian, where spin variables $s_i \in \{-1, +1\}$ are subject to local fields $h_i$ and pairwise interactions with coupling strengths $J_{ij}$.

Quantum computers use quantum bits (qubits) to hold information. Each qubit's behavior is governed by the laws of quantum mechanics, enabling qubits to be in a "superposition" state – that is, both a "-1" and a "+1" at the same time, until an outside event causes it to collapse into either a "-1" or a "+1" state. The output of an anneal is a low-energy ground state $\mathbf{s}$, which consists of an Ising spin for each qubit where $s_i \in \{-1, +1\}$. This is the basis upon which a quantum computer is constructed which gives the ability to quickly solve certain classes of NP-hard complex problems such as optimization, machine learning and sampling problems.

On the D-Wave device the connectivity between the binary variables $s_i$ is described by a fixed sparse graph $G = (V, E)$ called the *Chimera graph*. Nodes in $V$ as qubits represent problem variables with programmable weights, and edges as couplers in $E$ have programmable connection strengths. There are weights ($h_i$) associated with each qubit ($s_i$) and strengths ($J_{ij}$) associated with each coupler between qubits ($s_i$ and $s_j$). A quantum machine instruction (QMI) solves the objective function given the weights, strengths, and qubits. The D-Wave 2X system has 1095 qubits and 3061 couplers with sparse bipartite connectivity. The *Chimera graph* of this particular machine consists of a 12 x 12 array of 4 x 4 bipartite unit cells.

Often the quantum unconstrained binary optimization (QUBO) representation with it's 0/1-valued variables is more natural than the Ising -1/+1-valued variables. The QUBO objective function is shown in Eq. 2, where $Q$ is an $n \times n$ upper-triangular matrix of coupler strengths and $x$ is a vector of binary variables (0/1). $Q_{ii}$ is an analog to the Ising $h_i$, as are $Q_{ij}$ and $J_{ij}$. The Ising and QUBO models are related through the transformation $s = 2x - 1$. The D-Wave machine allows for either form.

$$O(\mathbf{Q}, \mathbf{x}) = \sum_i Q_{ii} x_i + \sum_{i<j} Q_{ij} x_i x_j \tag{2}$$

Physical constraints on current D-Wave platforms such as limited precision/control error and range on weights and strengths, sparse connectivity, and number of available qubits have an impact on the problem size and performance. Embedding algorithms are required to map or fit a problem graph onto the hardware. Strictly quantum approaches are limited by the number of graph nodes that can be represented on the hardware. Larger graphs require hybrid classical-quantum approaches.

The field of mathematics devoted to methods for efficiently partitioning a graph dates back to the 1970's [7, 8, 9]. Graph partitioning (GP) methods emerged to reduce the complexity of graphs for many different purposes such as applying divide and conquer techniques for efficient computation [10]. Other applications of GP include physical network design, VLSI design, telephone network design, load balancing of high performance computing (HPC) codes to minimize total communication between processors [11], distributed sparse matrix-vector multiplication (partitioning the rows of a matrix to minimize communication), physics lattices [12, 13], chemical elements related through bonds [14], metabolic networks [15] and social networks [16, 17, 18, 19].

Graph theory algorithms are used to determine the embedding of a problem graph on the D-Wave system using graph minors [20]. Other graph theory algorithms that have been implemented on the D-Wave system include graph coloring [21, 22], a graph isomorphism solver [23], and spanning tree calculations [24].

This work is motivated by a recently proposed graph-based electronic structure theory applied to quantum molecular dynamics (QMD) simulations [25]. In this approach, GP is used for reducing the calculation of the density matrix into smaller subsystems rendering the calculation computationally more efficient. This procedure is done at each timestep of a QMD simulation taking the previous density matrix as an adjacency matrix for the new graph.

When GP is unconstrained (with no limitation on partition size) and communication volume is minimized, the resulting natural parts are called communities. Community detection (CD) gives a high level "skeleton" view of the general structure of a graph based on a modularity metric [15, 19]. Detection and characterization of community structure in networks has been used to identify secondary structures in proteins [25] and to better understand the complex processes occurring in amino acid networks such as the allosteric mechanism in proteins [26].

Uniform or constrained GP partitions a graph into similar-sized parts while minimizing the number of *cut edges* between parts. Classical approaches to GP rely on heuristics and approximation algorithms. In this work, we initially address partitioning into 2 parts. Further partitioning into $2^N$ parts is a natural extension using recursive bisection. Better and faster results can be achieved with global methods, such as multi-level approaches. Multi-level implementations operate in stages, with refinement, as in METIS [27] and KaHIP [28]. Iterative multi-level GP can converge very quickly when the QUBOs used in the refinement are of sufficient size.

Considering the D-Wave architecture, a quantum GP approach should be able to partition a graph into $k$ parts concurrently, without recursion or stages. In our work, we were able to accomplish this by using ideas from the graph coloring problem [21, 22, 29]. A super-node is used to represent $k$ subnodes where the graph is split into $k$ parts. This is a natural formulation for the D-Wave computer, but quickly runs out of real estate for large graphs and large partitionings, requiring the use of a hybrid classical-quantum approach.

The D-Wave 2X computer is able to solve graphs of limited size (∼45 nodes). Graphs addressing interesting scientific questions quickly exceed this size. In order to extend this limit we explore techniques such as representing the graph by its complement for GP and reducing the coupler count by thresholding the modularity matrix for CD. Additionally, hybrid classical-quantum approaches can be employed, such as *qbsolv* [30], where processing on the CPU creates subgraphs to be run on the quantum processing unit (QPU) and assembled for the final result.

Following, we describe our methods for traditional algorithms for CD and GP implemented on the D-Wave quantum annealer, as well as methods formulated to take advantage of the hardware. Results are compared with existing benchmarks and current "state of the art" tools.

## 2. Methods

### 2.1. Graph Clustering/Community Detection

A graph can be divided in sets of nodes belonging to different communities (also called clusters). Nodes within any of these communities have a high probability of being connected (high intraconnectivity); whereas nodes in different communities have a lower probability of being connected (low interconnectivity) (see Fig. 1a). This natural division of a graph into communities differs from the usual GP problem in that the size of the communities cannot be predefined a priori.

A recently proposed metric that can quantify the quality of a community structure is the modularity [31, 19]. This metric performs a comparison of the connectivity of edges within communities with the connectivity of an equivalent network where edges are placed randomly. Let $G = (V, E)$ be a weighted graph with nodes $i$ in $V$ and edges $(i, j)$ in $E$ such that the corresponding adjacency matrix $A$ is defined as follows:

$$A_{ij} = \begin{cases} 0, & \text{if } i = j \\ w_{ij}, & \text{if } i \neq j \end{cases} \tag{3}$$

with $w_{ij}$ being the weight of edge $(i, j)$. The modularity matrix $B$ is taken as the difference between $A$ and a matrix constructed as an outer product of the vector degree **g**. The node degree $g_i$ is defined as $g_i = \sum_j A_{ij}$. Newman's expression for the modularity matrix $B$ can be rewritten as follows:

$$B = A - \frac{\mathbf{g}\mathbf{g}^T}{2m} \tag{4}$$

Or equivalently:

$$B_{ij} = A_{ij} - \frac{g_i g_j}{2m} = A_{ij} - \frac{g_i g_j}{\sum_l g_l} \tag{5}$$

If we now have a vector of labels $\mathbf{s}$, with $s_i \in \{-1, +1\}$ with "-1" or "+1" classifying nodes corresponding to different communities, the problem of finding the optimal modularity requires solving $\max_{\mathbf{s}}(Q(\mathbf{s}))$, or $\min_{\mathbf{s}}(-Q(\mathbf{s}))$ where:

$$Q(\mathbf{s}) = \mathbf{s}^T B \mathbf{s} \qquad (6)$$

By explicitly writing all the terms of Eq. 6, we can identify that the coupler strength $J_{ij}$ and onsite energy $h_i$ will have to be set to $-2B_{ij}$ and $-B_{ii}$ respectively to embed the problem on the D-Wave system. We can see that matrix $B$ does not impose any restrictions and requires no reformulation for the embedding. The problem of Eq. 6 is formulated as an Ising problem and as a consequence it is straightforward to apply quantum annealing techniques using the D-Wave system. In order to divide the graph into $2^N$ communities, a recursive subdivision can be performed where each of the communities is subdivided into two new communities.

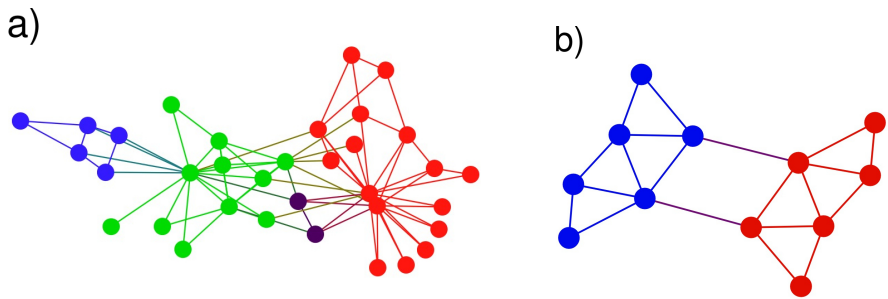

Figure 1: Example of community detection (a) showing four different communities in blue, green, purple and red colors containing 5, 11, 2 and 16 nodes respectively; and example of graph partitioning (b) showing two parts of equal size (six nodes) depicted with red and blue colors.

## 2.2. The Graph Partitioning Problem

Across multiple disciplines, graphs are frequently used to model different application problems. The sizes of these graphs can be arbitrarily large compared to the computational resources at hand. In order to reduce the complexity or enable parallelization, regardless of the application, a common technique used is to partition the graph into smaller subproblems. It has been shown that GP is an NP-hard problem thus achieving an efficient exact solution seems unlikely unless $P = NP$ [32, 33].

Popular GP algorithms such as Kernighan-lin [9], Fiduccia-Mattheyses [34] algorithms, or spectral methods incorporated into multi-level methods are often used. Since GP is often studied as a combinatorial optimization problem, combinatorial optimization methods, in particular, metaheuristics such as simulated annealing are also commonly used. In this work, we demonstrate how quantum annealing, a metaheuristic, can be used for partitioning a graph. We first formally define the GP problem and then relate it to quantum annealing and show how it can be mapped onto the D-Wave hardware.

Let $G = (V, E)$ be a graph with vertex set $V$ and edge set $E$ such that $n = |V|$ (number of vertices) and $m = |E|$ (number of edges). For a fixed integer $k$, the GP problem is to find a partition, $\Pi = (\Pi_1, \ldots, \Pi_k)$, of the vertex set $V$ into $k$ equal parts such that the number of *cut edges* is minimized. A *cut edge* is defined as an edge whose end points are in different partitions. In order to map the GP problem on the D-Wave hardware, we need to formulate it as a QMI. We first formulate it for $k = 2$ and later generalize it for arbitrary $k$. For $k = 2$, the objective of minimizing the number of *cut edges* can be formulated into a quadratic programming problem as follows: Label each vertex $i$ with $s_i = \pm 1$, depending on the part they belong to, then the number of *cut edges* is given by:

$$N_c = \frac{1}{4} \sum_{(i,j) \in E} (s_i - s_j)^2 \qquad (7)$$

subject to the balancing constraint

$$\sum_i s_i = 0. \qquad (8)$$

Let $\tilde{G}$ be a fixed orientation of $G$ (where each edge can have any assigned direction). Then, the *incidence* matrix of $\tilde{G}$ is a $n \times m$ matrix $C$ defined as

$$C_{il} = \begin{cases} 1, & e_l = (i, k) \\ -1, & e_l = (k, i) \\ 0, & \text{otherwise.} \end{cases} \qquad (9)$$

where $e_l = (i, k)$ and $e_l = (k, i)$ indicate edges with $i \to k$ and $k \to i$ directions respectively; and $l$ runs through all the edges connecting node $i$. If $\mathbf{s}$ is a vector of vertex labels then $\mathbf{s}^T C$ is a $1 \times m$ vector with entry $(s_i - s_j)$ for each edge. Therefore,

$$||\mathbf{s}^T C||^2 = \sum_{(i,j) \in E} (s_i - s_j)^2 \qquad (10)$$

.

The *Laplacian* matrix, $L$, of $G$ is defined by

$$L = D - A \qquad (11)$$

where $D$ is a diagonal matrix with entry $D_{ii}$ equal to the degree $g_i$ of vertex $i$. $D_{ii} = g_i = \sum_j A_{ij}$) and $A$ is the adjacency matrix [35]. One can easily show that matrix $L$ is equivalent to:

$$L = CC^T \qquad (12)$$

So $\mathbf{s}^T L \mathbf{s} = \mathbf{s}^T C C^T \mathbf{s} = ||\mathbf{s}^T C||$. Thus, the graph partitioning problem is formulated as:

$$\min_{\mathbf{s}}(\tfrac{1}{4}\mathbf{s}^T L \mathbf{s})$$

subject to $\sum_i s_i = 0$

with $\quad s_i \in \{-1, 1\}, \ i = 1, ..., n$ \hfill (13)

and equivalently

$$\min_{\mathbf{x}}(\mathbf{x}^T L \mathbf{x})$$

subject to $\sum_i x_i = \dfrac{n}{2}$

with $\quad x_i \in \{0, 1\}, \quad i = 1, ..., n$ \hfill (14)

Another equivalent formulation for GP is

$$\max_{\mathbf{s}}(\mathbf{s}^T A \mathbf{s})$$

subject to $\sum_i s_i = 0$

with $\quad s_i \in \{-1, 1\}, \ i = 1, ..., n$ \hfill (15)

whose objective function is the difference between the number of edges and twice the number of *cut edges*.

In order to partition a graph using a quantum annealer, the graph partitioning problem must be reformulated as a QUBO or Ising problem in analogy with Eq. 1 or 2. In order to do this, the constraints in the formulation (15) need to be removed, leading to a relaxation of the original problem as follows:

$$\max_{\mathbf{s}}(\beta \mathbf{s}^T A \mathbf{s} - \alpha (\textstyle\sum s_i)^2)$$

with $\ s_i \in \{-1, 1\} \ i = 1, ..., n$ \hfill (16)

where $\alpha$ and $\beta$ are weight parameters such that an increase in $\alpha$ indicates an importance of the balancing criterion while an increase in $\beta$ indicates an importance for a smaller cut over the balancing criterion. $\alpha$ and $\beta$ are chosen according to [36]. We refer to formulation (16) as the Ising formulation where variable $s \in \{-1, +1\}$.

Since

$$(\textstyle\sum s_i)^2 = \sum s_i^2 + 2 \sum s_i s_j = \mathbf{s}^T \mathbb{1}_{n \times n} \mathbf{s}, \hfill (17)$$

where $\mathbb{1}_{n \times n}$ is the $n \times n$ matrix with all entries equal to 1, we can rewrite formulation (16) as

$$\max_{\mathbf{s}}(\mathbf{s}^T(\beta A - \alpha \mathbb{1}_{n \times n})\mathbf{s})$$

with $\ s_i \in \{-1, 1\} \ i = 1, ..., n$ \hfill (18)

Or equivalently,

$$\min_{\mathbf{s}}(\mathbf{s}^T(\alpha \mathbb{1}_{n \times n} - \beta A)\mathbf{s})$$

with $\ s_i \in \{-1, 1\} \ i = 1, ..., n$ \hfill (19)

In order to transform the Ising model into a QUBO, we use the transformation

$$\mathbf{s} = 2\mathbf{x} - \mathbb{1}_n \tag{20}$$

where $\mathbb{1}_n$ is the vector of all ones and $\mathbf{x} \in \{0,1\}^n$. In general, if we apply the transformation to any symmetric matrix $M$, a straightforward substitution and simplification gives the transformation

$$\mathbf{s}^T M \mathbf{s} = 4\mathbf{x}^T M \mathbf{x} - 4\mathbf{x}^T M \mathbb{1}_n + \mathbb{1}_n^T M \mathbb{1}_n. \tag{21}$$

Thus, taking $M = \alpha \mathbb{1}_{n \times n} - \beta A$, we have

$$\begin{aligned}
&\min_{\mathbf{s}}(\mathbf{s}^T(\alpha \mathbb{1}_{n \times n} - \beta A)\mathbf{s}) \\
&= \min_{\mathbf{x}}(\ \mathbf{x}^T(\alpha \mathbb{1}_{n \times n} - \beta A)\mathbf{x} - \mathbf{x}^T(\alpha \mathbb{1}_{n \times n} - \beta A)\mathbb{1}_n)
\end{aligned} \tag{22}$$

Now let $Q_{ij}$ be the coefficients of the QUBO, then

$$Q_{ij} = \begin{cases} \alpha - \beta, & \text{if } (i,j) \in E \\ \alpha, & \text{if } (i,j) \notin E, i \neq j \end{cases} \tag{23}$$

and $Q_{ii} = \beta\, g_i - \alpha(n-1)$. Here, $g_i$ is the degree of vertex $i$. The above solution follows because:

$$\mathbf{x}^T(\alpha \mathbb{1}_{n \times n} - \beta A)\mathbb{1}_n = \mathbf{x}^T \begin{bmatrix} \alpha n - \beta\, g_1 \\ \alpha n - \beta\, g_2 \\ \vdots \\ \alpha n - \beta\, g_n \end{bmatrix} = -\mathbf{x}^T \begin{bmatrix} \beta\, g_1 - \alpha n \\ \beta\, g_2 - \alpha n \\ \vdots \\ \beta\, g_n - \alpha n \end{bmatrix}$$

since $A\mathbb{1}_n$ is the vector whose $i$th entry is $g_i$ and also taking the diagonal elements of the first part of (22) into account. These QUBO variables are then mapped onto the *Chimera graph* of the D-Wave machine to perform the calculation. Because of physical limitations of the hardware, embedding algorithms must be used. Note that when $\alpha = \beta$, the graph complement is mapped onto the *Chimera graph*. This is particularly helpful for dense graphs in reducing the number of qubits and couplers required for the embedding.

## 2.3. k-Concurrent Approach for Graph Partitioning

The $k$-Concurrent approach allows us to partition a graph into $k$ parts in parallel without recursion. This is a general approach that can be applied to clustering or partitioning. Similar to the graph coloring problem [21, 22, 29] each graph vertex is represented by a super-node consisting of $k$ subnodes, where $k$ is the number of partitions desired (see Fig. 2). Each of the $k$ subnodes has a unary encoding (either "0" or "1"). After GP, only one of the subnodes is set to "1", while the rest are "0" for each vertex denoting which part it belongs to.

Adjacent vertices are given different colors (or parts) in the graph coloring problem. In constrast, partitioning influences adjacent vertices to be in the same part. Super-nodes are connected by super-edges, pairing corresponding subnodes. Within each super-node, subnodes are constrained such that only one will be turned on as in the graph coloring problem [21, 22]. Using $k$-Concurrent GP results in $k$ copies of the graph in parallel as a $kN \times kN$ matrix to represent the partitioning problem.

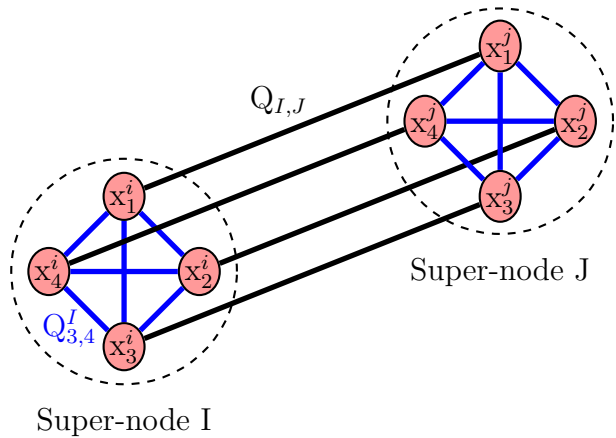

Figure 2: An example of the super-node concept used in $k$-Concurrent GP is shown for partitioning into 4 parts. Two super-nodes $I$ and $J$ consisting of four subnodes each are connected by a super-edge $Q_{I,J}$. Internal edges $Q_{l,m}^{I/J}$ where $l, m \in \{1 - 4\}$ are set to enforce the selection of only one subnode to be equal to "1" after GP. The super-edge $Q_{I,J}$ is shown with connections between corresponding subnodes.

In order to partition a graph concurrently into an arbitrary number of parts, $k$, we need to formulate the problem as a QUBO, which takes on binary variables. The authors in [37] provide different mathematical formulations for graph partitioning. In this section, we make a slight modification to the standard formulation and reformulate it as a quadratic program, which generalizes the above formulation of partitioning into 2 parts. We then relax the mathematical formulation as a QUBO written in matrix form.

The decision variables are given by

$$x_{i,j} = \begin{cases} 1, & \text{if node } i \text{ is in part } j \\ 0, & \text{otherwise.} \end{cases}$$

The constraint

$$\sum_{j=1}^{k} x_{i,j} = 1 \tag{24}$$

for each node $i$ ensures that each node is in exactly one part. While

$$\sum_{i=1}^{n} x_{i,j} = \frac{n}{k}$$

for $j = 1, \ldots, k$ are the balancing constraints for the part sizes.

$$\mathbf{x}_j = \begin{bmatrix} x_{1,j} \\ x_{2,j} \\ \vdots \\ x_{n,j} \end{bmatrix},$$

The number of cut edges across the $k$ parts is given by

$$\frac{1}{2} \left( \sum_{j=1}^{k} \mathbf{x}_j^T L \mathbf{x}_j \right) \tag{25}$$

where $L$ is the Laplacian matrix. Thus, giving us a generalization

$$\min_{\mathbf{x}} \left( \sum_{j=1}^{k} \mathbf{x}_j^T L \mathbf{x}_j \right)$$

$$\text{subject to} \quad \sum_{i=1}^{n} x_{i,j} = \frac{n}{k} \qquad j = 1, \ldots, k$$

$$\sum_{j=1}^{k} x_{i,j} = 1 \qquad i = 1, \ldots, n \tag{26}$$

$$\text{with} \qquad x_{i,j} \in \{0, 1\}, \qquad i = 1, \ldots, n. \ j = 1, \ldots, k$$

A relaxation of the above formulation is given by

$$\beta \left( \sum_{j=1}^{k} \mathbf{x}_j^T L \mathbf{x}_j \right) + \sum_{j=1}^{k} \alpha_j \left( \sum_{i=1}^{n} x_{i,j} - \frac{n}{k} \right)^2 + \sum_{i=1}^{n} \gamma_i \left( \sum_{j=1}^{k} x_{i,j} - 1 \right)^2 \tag{27}$$

where $\beta, \alpha_i$ and $\gamma_i$ are positive penalty contants.

Let $\mathcal{N} = nk$, $\mathbf{X}^T = [\mathbf{x}_1^T \ \mathbf{x}_2^T \ \ldots \ \mathbf{x}_k^T]$. Define $\mathcal{L}$ as a block diagonal matrix with $k$ copies of the Laplacian matrix on the diagonal and $\alpha \mathcal{I}$ as block matrix, with $k \times k$ blocks, given by

$$\alpha \mathcal{I} = \begin{pmatrix} \alpha_1 \mathbb{1}_{n \times n} & 0 & \ldots & & 0 \\ 0 & \alpha_2 \mathbb{1}_{n \times n} & \ldots & & \vdots \\ 0 & 0 & \ddots & & \vdots \\ \vdots & \vdots & \ddots & \ddots & 0 \\ 0 & 0 & \ldots & 0 & \alpha_k \mathbb{1}_{n \times n} \end{pmatrix}$$

and the block vector $\alpha \mathbb{1}_{\mathcal{N}}$ and $\Gamma$ as

$$\alpha \mathbb{1}_{\mathcal{N}}^T = \begin{pmatrix} \alpha_1 \mathbb{1}_n^T & \alpha_2 \mathbb{1}_n^T & \ldots & \alpha_k \mathbb{1}_n^T \end{pmatrix}$$

$$\Gamma^T = \underbrace{\begin{pmatrix} \gamma^T & \gamma^T & \ldots & \gamma^T \end{pmatrix}}_{k \text{ times}}$$

where $\gamma^T = \begin{pmatrix} \gamma_1 & \gamma_2 & \dots & \gamma_n \end{pmatrix}^T$. If $\mathbf{I}_m$ is an identity matrix of size $m$, then the terms in (27) can be written in matrix form as:

$$\beta \sum_{j=1}^{k} \mathbf{x}_j^T L \mathbf{x}_j = \beta \mathbf{X}^T \mathcal{L} \mathbf{X},$$

with

$$\sum_{j=1}^{k} \alpha_j (\sum_{i=1}^{n} x_{i,j} - \frac{n}{k})^2 = \sum_{j=1}^{k} \alpha_j \Big( (\sum_{i=1}^{n} x_{i,j})^2 - 2\frac{n}{k} \sum_{i=1}^{n} x_{i,j} + \frac{n^2}{k^2} \Big)$$

$$= \sum_{j=1}^{k} \alpha_j \Big( \mathbf{x}_j^T \mathbb{1}_{n \times n} \mathbf{x}_j - 2\frac{n}{k} \mathbb{1}_n^T \mathbf{x}_j + \frac{n^2}{k^2} \Big)$$

$$= \mathbf{X}^T \alpha \mathcal{I} \mathbf{X} - 2\frac{n}{k} \alpha \mathbb{1}_{\mathcal{N}}^T \mathbf{X} + \frac{n^2}{k^2} \sum_{j=1}^{k} \alpha_j,$$

and

$$(\sum_{j=1}^{k} x_{i,j} - 1)^2 = (\sum_{j=1}^{k} x_{i,j})^2 - 2\sum_{j=1}^{k} x_{i,j} + 1$$

Let $\mathbf{Z}_i$ the $\mathcal{N} \times \mathcal{N}$ zero matrix whose $j$th diagonal element is 1 if and only if $j \equiv i \pmod{n}$. For example in $\mathbf{Z}_1$, every $1^{\text{st}}, (n+1)^{\text{th}}, (2n+1)^{\text{th}}, \dots, ((k-1)n+1)^{\text{th}}$ diagonal element is 1 and has zero everywhere else. Then

$$\Big( \sum_{j=1}^{k} x_{i,j} \Big)^2 = \mathbf{X}^T \mathbf{Z}_i \mathbb{1}_{\mathcal{N} \times \mathcal{N}} \mathbf{Z}_i \mathbf{X}$$

and

$$\sum_{j=1}^{k} x_{i,j} = \mathbb{1}_{\mathcal{N}}^T \mathbf{Z}_i \mathbf{X}.$$

So,

$$(\sum_{j=1}^{k} x_{i,j} - 1)^2 = \mathbf{X}^T \mathbf{Z}_i \mathbb{1}_{\mathcal{N} \times \mathcal{N}} \mathbf{Z}_i \mathbf{X} - 2\mathbb{1}_{\mathcal{N}}^T \mathbf{Z}_i \mathbf{X} + 1$$

Since,

$$\sum_{j=1}^{k} \alpha_j (\sum_{i=1}^{n} x_{i,j} - \frac{n}{k})^2 = \mathbf{X}^T \alpha \mathcal{I} \mathbf{X} - 2\frac{n}{k} \alpha \mathbb{1}_{\mathcal{N}}^T \mathbf{X} + \frac{n^2}{k^2} \sum_{j=1}^{k} \alpha_j,$$

and

$$\sum_{i=1}^{n} \gamma_i (\sum_{j=1}^{k} x_{i,j} - 1)^2 = \sum_{i=1}^{n} \gamma_i \Big( \mathbf{X}^T \mathbf{Z}_i \mathbb{1}_{\mathcal{N} \times \mathcal{N}} \mathbf{Z}_i \mathbf{X} - 2\mathbb{1}_{\mathcal{N}}^T \mathbf{Z}_i \mathbf{X} + 1 \Big)$$

$$= \mathbf{X}^T \sum_{i=1}^{n} \gamma_i \Big( \mathbf{Z}_i \mathbb{1}_{\mathcal{N} \times \mathcal{N}} \mathbf{Z}_i \Big) \mathbf{X} - 2\sum_{i=1}^{n} \gamma_i \mathbb{1}_{\mathcal{N}}^T \mathbf{Z}_i \mathbf{X} + \sum_{i=1}^{n} \gamma_i.$$

Let $D_\gamma$ be a diagonal matrix such that

$$D_\gamma = \text{diag}(\gamma_1, \ldots, \gamma_n)$$

and $\mathbf{B}_\Gamma$ be a block matrix with $k \times k$ blocks, where each block is equal to $D_\gamma$, then

$$\sum_{i=1}^n \gamma_i \mathbf{Z}_i \mathbb{1}_{\mathcal{N} \times \mathcal{N}} \mathbf{Z}_i = \mathbf{B}_\Gamma$$

and

$$\sum_{i=1}^n \gamma_i \mathbb{1}_{\mathcal{N}}^T \mathbf{Z}_i \mathbf{X} = \Gamma^T \mathbf{X}.$$

So,

$$\sum_{i=1}^n \gamma_i (\sum_{j=1}^k x_{i,j} - 1)^2 = \mathbf{X}^T \mathbf{B}_\Gamma \mathbf{X} - 2\Gamma^T \mathbf{X} + \sum_{i=1}^n \gamma_i.$$

So if $\mathcal{L}$ is the block diagonal matrix with the laplacian matrix of $G$ at each block, then the QUBO is equivalent to

$$\min_{\mathbf{x}} (\mathbf{X}^T (\beta \mathcal{L} + \alpha \mathcal{I} + \mathbf{B}_\Gamma) \mathbf{X} - (2\Gamma^T + 2\frac{n}{k} \alpha \mathbb{1}_{\mathcal{N}}^T) \mathbf{X})$$

$$\text{with} \quad x_{i,j} \in \{0, 1\}, \ i = 1, ..., n. \ j = 1, \ldots, k$$

(28)

## 3. Results and Discussion

Our GP experiments on the D-Wave machine used software tools such as *sapi* Python [38] for graphs that fit onto the architecture (up to $\sim 70$ verices) and the hybrid classical-quantum *qbsolv* [30] for larger graphs (up to $\sim$9000 vertices). NetworkX [39] was used for generating and processing graphs as part of this work. Random graph models (e.g. Erdos-Renyi, PowerLaw), large graphs from the Walshaw Archive [40, 41], and QMD molecule electronic structure graphs [42] were used to evaluate our methods. The quality of the GP was evaluated by a comparison metric as the number of *cut edges* between partitions (smaller is better). The results were compared to existing multi-level GP frameworks, METIS [27] and KaHIP [28] (winner of the 10th DIMACS challenge), or in some cases the best known solution from the Walshaw Archive [41].

When running experiments using *sapi* directly on the D-Wave computer, multiple solutions are returned as a histogram, starting with the lowest energy. GP experiments using *qbsolv* return the lowest energy solution.

### 3.1. Community Detection with Thresholding

In Table 1 we show the results of modularity values for clustering into two communities using the D-Wave system. In this case we applied community detection to the karate club graph (34 vertices) and used *qbsolv* to solve the problem on the *chimera graph*.

We explored thresholding of the modularity matrix. In this case we can see that with a threshold value of 0.12 meaning that we set $B_{ij} = 0$ if $B_{ij} < 0.12$, modularity is reduced by $\sim 30$ % but the required number of edges is significantly less ($\sim 65$ % less) which has direct consequence on the number of qubits/couplers required for embedding on the D-Wave hardware.

Table 1: Graph 2-Clustering with Thresholding

| Threshold | # Edges | Modularity |
|:---:|:---:|:---:|
| 0 | 561 | 0.37179487 |
| 0.02 | 544 | 0.37179487 |
| 0.05 | 411 | 0.37146614 |
| 0.07 | 300 | 0.37146614 |
| 0.08 | 244 | 0.27714497 |
| 0.10 | 227 | 0.27714497 |
| 0.11 | 212 | 0.25509533 |
| 0.12 | 194 | 0.25509533 |
| 0.13 | 169 | 0 |

In Table 2 we show results for 4-clustering using the recursive method. Random graphs with powerlaw degree distribution (powerlaw cluster graph generator with 100 vertices) were used. In this case thresholding is also possible but the quality of the clustering is significantly reduced.

Table 2: Recursive Graph 4-Clustering with Thresholding

| Threshold | # Edges | Modularity |
|:---:|:---:|:---:|
| 0 | 4950 | 0.22655990 |
| 0.06 | 4593 | 0.22651496 |
| 0.08 | 3799 | 0.18783171 |
| 0.10 | 3091 | 0.06679333 |

The reduction in the number of edges in both cases will result in a reduction of the number of qubits/couplers that are needed to embed the matrix into the QPU of the D-Wave machine. Preprocessing of the modularity matrix to produce a sparse version that is still representative of the original allows for larger problems to be solved efficiently by community clustering using quantum annealing.

## 3.2. Graph Partitioning

For partitioning a graph into 2 parts of equal or similar size, we studied two sets of graphs chosen in order to determine the limits of the different partitioning methods used. The first set contained graphs of a relatively small size, having at most 70 vertices. The second set consisted of graphs with at most 9000 vertices chosen from the graph partitioning archive by Chris Walshaw [40, 41], an online archive dedicated to documenting the best quality solutions from a set of benchmark graphs.

For the first set, we partitioned the graphs using *sapi* as our main interface to the D-Wave machine. Thus, 100 percent of the partitioning was carried out by the quantum annealer. We generated random graphs (Erdos-Renyi) with a variable probability parameter $p$ (fraction of total possible edges). The largest complete graph that is fully embeddable on the D-Wave 2X has approximately 45 vertices. However, our results demonstrate that for large values of $p$, (i.e., dense graphs) we can successfully partition such graphs even if they have more than 45 vertices. This was possible due to our use of the graph complement in the QUBO formulation. In our experiments, we partitioned graphs with up to 70 vertices. All our results gave solutions with a comparable quality to the solvers METIS and KaHIP. Table 3 shows the results for this first set of graphs.

Table 3: Graph 2-Partitioning using *sapi*

| N | METIS | KaHIP | *sapi* |
|---|---|---|---|
| simulator | | | |
| 20 | 82 | 82 | 82 |
| 30 | 183 | 182 | 182 |
| 40 | 326 | 324 | 330 |
| D-Wave 2X | | | |
| 40 | 334 | 334 | 334 |
| 60 | 766 | 765 | 768 |
| 70 | 1039 | 1042 | 1045 |

For the second set, we partitioned the graphs using *qbsolv* as our main interface to the D-Wave machine. Since *qbsolv* is a hybrid classical-quantum approach, not all the processing was carried out by the quantum annealer. This enabled us to partition graphs with over 100 vertices. Our solutions, shown in Fig. 4, gave high quality partitions, in all cases having a cut size smaller than the solutions from the two solvers, METIS and KaHIP. Our solutions matched the best known for the last 2 graphs.

Table 4: Graph 2-Partitioning using *qbsolv*

| Graph | N | Best | METIS | KaHIP | *qbsolv* |
|---|---|---|---|---|---|
| add20 | 2395 | 596 | 723 | 760 | 647 |
| data | 2851 | 189 | 225 | 221 | 191 |
| 3elt | 4720 | 90 | 91 | 92 | 90 |
| bcsstk33 | 8738 | 10171 | 10244 | 10175 | 10171 |

For partitioning a graph into $k = 2^i$ parts for $i > 1$, we performed our experiments on molecule electronic structure graphs as density matrices generated from QMD simulations [42]. As an example, Fig. 3 shows the phenyl dendrimer molecule composed of 22 covalently bonded phenyl groups with C and H atoms only with a total of 262 atoms. The electronic structure of this molecule consists of 730 orbitals, resulting in a graph of 730 vertices. This molecule is a special case, where the associated graph has a fractal-like structure and represents a difficult case for GP.

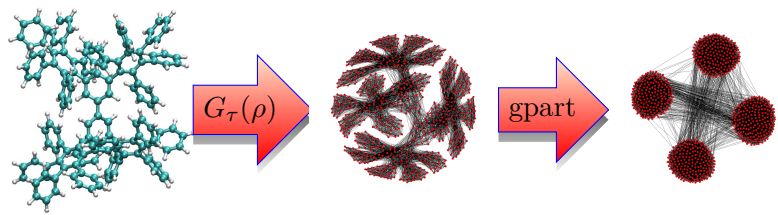

Figure 3: From left to right: Phenyl dendrimer molecular structure composed of 22 covalently bonded phenyl groups with C and H atoms only shown in cyan and white colors respectively. This molecule consists of 262 atoms with 730 orbitals. We also show the resulting electronic structure graph $G_\tau(\rho)$ constructed from the QMD density matrix $\rho$ using a threshold $\tau$ that leads to 730 vertices (see reference [42]); and the resulting 4-Concurrent GP using the D-Wave machine where nodes are grouped by which part they belong to.

Table 5: Recursive Graph k-Partitioning of molecule electronic structure graphs using *qbsolv*

| k | METIS | KaHIP | *qbsolv* |
|---|---|---|---|
| Phenyl dendrimer N=730 | | | |
| 2 | 705 | 705 | 705 |
| 4 | 2636 | 2638 | 2654 |
| 8 | 12621 | 12677 | 15769 |
| Peptide 1aft N=384 | | | |
| 2 | 3 | 6 | 3 |
| 4 | 37 | 35 | 22 |
| 8 | 72 | 78 | 66 |

Partitioning each molecule electronic structure graph into 2 parts using *qbsolv* results in equal or reduced number of *edge cuts* compared to METIS and KaHIP as seen in Table 5. Recursive bisection into 4 and 8 parts results in comparable solutions, though increased number of *edge cuts* for the Phenyl dendrimer. Better results are seen for the Peptide 1aft protein compared to METIS and KaHIP.

### 3.3. k-Concurrent Graph Partitioning

*k*-Concurrent GP requires that a graph of size $N$ (number of nodes) $\times$ $k$ (number of partitions) be embedded in the *Chimera graph*. This quickly uses up the available qubits. Results of running concurrent GP on small random graphs on the D-Wave QPU only using *sapi* are shown in Table 6. The number of *cut edges* between partitions are shown. Concurrent GP been formulated to produce equal sized partitions. Partition sizes never differ by more than one. In this case, METIS comparisons are contrained to equal sized partitions or very close (using the -ufactor=1 option). Results using *sapi* are comparable to METIS and *qbsolv*. Running directly on the D-Wave machine using *sapi* limits the the embeddable graph size to ~45 nodes. A 15 node graph partitioned into 4 parts used almost all available qubits. Similarly for 20 nodes split into 3 parts. We can see that running *k*-Concurrent GP on the QPU produces quality results.

Table 6: *k*-Concurrent Graph Partitioning using *sapi*. We show the number of *cut edges* between partitions.

| N | k | *sapi* | METIS | *qbsolv* |
|---|---|--------|-------|----------|
| 10 | 2 | 19 | 19 | 19 |
| 10 | 3 | 29 | 29 | 29 |
| 10 | 4 | 32 | 33 | 32 |
| 11 | 2 | 25 | 26 | 25 |
| 11 | 3 | 36 | 36 | 36 |
| 11 | 4 | 38 | 39 | 38 |
| 15 | 2 | 45 | 47 | 45 |
| 15 | 3 | 62 | 62 | 62 |
| 15 | 4 | 70 | 73 | 70 |
| 20 | 2 | 83 | 83 | 83 |
| 20 | 3 | 120 | 122 | 120 |
| 23 | 2 | 109 | 114 | 109 |
| 27 | 2 | 156 | 164 | 156 |
| 30 | 2 | 182 | 183 | 182 |

*k*-Concurrent GP on large graphs requires the use of the hybrid classical-quantum *qbsolv*. Results for random dense graphs of 250, 500, and 1000 nodes split into 2, 4, 8, and 16 parts are shown in Table 7. The quality of the partitionings is comparable to METIS, while the number of *cut edges* is consistently reduced by tens to hundreds.

In Table 8 the results for *k*-Concurrent GP of the molecule electronic structure graphs are shown. Fig. 3 shows the molecular structure for the Phenyl dendrimer, followed by the electronic structure graph from the QMD density matrix, and the resulting 4-Concurrent GP. Due to the fractal structure, this is a difficult case for GP. The *qbsolv* results are comparable or better than METIS for the Phenyl dendrimer and the Peptide 1aft. We see a very large number of *cut edges* for METIS when 4-partitioning the Phenyl Dendrimer due to the constraint on equal partitions.

Table 7: *k*-Concurrent Graph Partitioning using *qbsolv*

| N | k | METIS | *qbsolv* |
|---|---|-------|----------|
| 250 | 2 | 13691 | 13600 |
| | 4 | 20884 | 20587 |
| | 8 | 24384 | 24459 |
| | 16 | 26224 | 26176 |
| 500 | 2 | 55333 | 54999 |
| | 4 | 83175 | 83055 |
| | 8 | 98073 | 97695 |
| | 16 | 105061 | 105057 |
| 1000 | 2 | 221826 | 221420 |
| | 4 | 334631 | 334301 |
| | 8 | 392018 | 392258 |
| | 16 | 421327 | 420970 |

Table 8: *k*-Concurrent Graph Partitioning of molecule electronic structure graphs using *qbsolv*

| k | METIS | *qbsolv* |
|---|-------|----------|
| Phenyl dendrimer N=730 | | |
| 2 | 706 | 706 |
| 4 | 20876 | 2648 |
| 8 | 22371 | 15922 |
| 16 | 28666 | 26003 |
| Peptide 1aft N=384 | | |
| 2 | 12 | 12 |
| 4 | 29 | 20 |
| 8 | 121 | 66 |
| 16 | 209 | 180 |

## 4. Conclusion

We have shown that GP framed as a QUBO problem is a natural fit for the D-Wave quantum annealer. Our results using quantum and hybrid classical-quantum approaches are comparable or better than existing traditional GP tools. We showed that solving CD using a thresholded modularity matrix did not change the community results and could be run in a reduced qubit/coupler footprint. 2-partitioning of the Walshaw archive graphs and random graphs equaled or outperformed existing GP tools and in some cases the best known results. A *k*-Concurrent GP approach using the super-node concept based on the map coloring problem was used to partition random graphs and molecule electron structure graphs into 2, 4, 8, and 16 parts, improving the quality of the partitioning over existing tools by reducing the number of *cut edges* between parts by tens to hundreds. *k*-Concurrent GP was shown to run on the QPU directly for small graphs and using hybrid classical-quantum *qbsolv* for large graphs.

Quantum annealing GP approaches were shown to produce quality partitions for example graphs as well as electronic structure graphs from QMD simulations. Future plans include applying *k*-Concurrent GP to other domains and extending the *k*-Concurrent approach to CD for discovering communities in graph structure problems.

## 5. Acknowledgements

The authors would like to acknowledge D-Wave Systems for their useful tutorials and use of the Burnaby D-Wave machine. The authors would also like to acknowledge the NNSA's Advanced Simulation and Computing (ASC) program at Los Alamos National Laboratory (LANL) for use of their Ising D-Wave 2X quantum computing resource. This research has been funded by the Los Alamos National Laboratory (LANL) Information Science and Technology Institute (ISTI) and Laboratory Directed Research and Development (LDRD). Assigned: Los Alamos Unclassified Report 17-23649. LANL is operated by Los Alamos National Security, LLC, for the National Nuclear Security Administration of the U.S. DOE under Contract DE-AC52-06NA25396.

## 6. References

[1] McGeoch C C 2014 Synthesis Lectures on Quantum Computing **5**(2) 1–93

[2] Boixo S, Rønnow T F, Isakov S V, Wang Z, Wecker D, Lidar D A, Martinis J M and Troyer M 2014 Nature Physics **10**(3) 218–224

[3] Los Alamos National Laboratory 2016 1663 Magazine 14–19

[4] Lanting T, Przybysz A J, Smirnov A Y, Spedalieri F M, Amin M H, Berkley A J, Harris R, Altomare F, Boixo S, Bunyk P, Dickson N, Enderud C, Hilton J P, Hoskinson E, Johnson M W, Ladizinsky E, Ladizinsky N, Neufeld R, Oh T, Perminov I, Rich C, Thom M C, Tolkacheva E, Uchaikin S, Wilson A B and Rose G 2014 Phys. Rev. X **4**(2) 021041

[5] Albash T, Rannow T, Troyer M and Lidar D 2015 The European Physical Journal Special Topics **224**(1) 111–120

[6] Boixo S, Smelyanskiy V N, Shabani A, Isakov S V, Dykman M, Denchev V S, Amin M H, Smirnov A Y, Mohseni M and Neven H 2016 Nature Communications **7** 10327

[7] Buluç A, Meyerhenke H, Safro I, Sanders P and Schulz C 2013 CoRR **abs/1311.3144**

[8] Sanders P and Schulz C 2012 High quality graph partitioning Graph Partitioning and Graph Clustering pp 1–18

[9] Kernighan B W and Lin S 1970 The Bell System Technical Journal **49** 291–307

[10] Kobayashi M and Nakai H 2011 Linear scaling techniques in computational chemistry vol 13 (Springer)

[11] Hendrickson B and Kolda T G 2000 Parallel Computing **26** 1519 – 1534 ISSN 0167-8191 graph Partitioning and Parallel Computing

[12] Haenggi M 2010 CoRR **abs/1004.0027**

[13] Fatt I 1956

[14] Randić M and Plavsić D 2003 International Journal of Quantum Chemistry **91** 20–31 ISSN 1097-461X

[15] Girvan M and Newman M E J 2002 Proc. Natl. Acad. Sci. **99** 7821–7826

[16] Grönlund A 2006 Complex patterns: from physical to social interactions Ph.D. thesis Umeå University

[17] Brin S and Page L 1998 Computer Networks ISDN **30** 107–117

[18] Solé R V and Valverde S 2004 Complex networks vol 650 (Springer)

[19] Newman M E J 2006 Proc. Natl. Acad. Sci. **103** 8577–8582

[20] Cai J, Macready W G and Roy A 2014 arXiv preprint arXiv:1406.2741

[21] Dahl D 2016 2016 LANL D-Wave Tutorials

[22] Dahl E D 2013 D-Wave Systems Whitepaper 1–12

[23] Zick K, Shehab O and Matthew F 2015 Sci. Rep. **5**

[24] Novotny M A, Hobl L and Hall J S 2016 Journal of Physics: Conference Series **681** 012005

[25] Niklasson A M N, Mniszewski S M, Negre C F A, Cawkwell M J, Swart P J, Mohd-Yusof J, Germann T C, Wall M E, Bock N, Rubensson E H and Djidjev H 2016 Journal of Chemical Physics **144** 234101

[26] Rivalta I, Sultan M M, Lee N S, Manley G A, Loria J P and Batista V S 2012 Proceedings of the National Academy of Sciences **109** E1428–E1436

[27] Karypis G and Kumar V 1999 SIAM Journal on Scientific Computing **20**(1) 359–392

[28] Sanders P and Schulz C 2013 Think locally, act globally: Highly balanced graph partitioning Proceedings of the 12th International Symposium on Experimental Algorithms pp 164–175

[29] Kochenberger G A, Glover F, Alidaee B and Rego C 2005 Ann Oper Res **139** 229–241

[30] Booth M, Reinhardt S P and Roy A 2017 D-Wave Technical Report Series **14** 1–9 URL http://github.com/dwavesystems/qbsolv

[31] Newman M E and Girvan M 2004 Physical review E **69** 026113

[32] Garey M R, Johnson D S and Stockmeyer L 1976 Theoretical computer science **1** 237–267

[33] Hyafil L and Rivest R L 1973 Graph partitioning and constructing optimal decision trees

are polynomial complete problems (IRIA. Laboratoire de Recherche en Informatique et Automatique)

[34] Fiduccia C M and Mattheyses R M 1988 A linear-time heuristic for improving network partitions Papers on Twenty-five years of electronic design automation (ACM) pp 241–247

[35] Jeribi A 2015 Spectral theory and applications of linear operators and block operator matrices (Springer)

[36] Lucas A 2013 arXiv preprint arXiv:1302.5843

[37] Boulle M 2004 Optimization and Engineering **5** 315–333

[38] D-Wave-Systems 2016 D-Wave Technical Report Series **09** 1–71

[39] Hagberg A A, Shult D A and Swart P J 2008 Exploring network structure, dynamics, and function using networkx Proc. SciPy 2008 pp 11–16 URL `http://networkx.github.io/`

[40] Soper A J, Walshaw C and Cross M 2004 Journal of Global Optimization **29** 225–241

[41] Walshaw C 2016 URL `http://chriswalshaw.co.uk/partition/`

[42] Djidjev H N, Hahn G, Mniszewski S M, Negre C F, Niklasson A M and Sardeshmukh V B 2016 Graph partitioning methods for fast parallel quantum molecular dynamics 2016 Proceedings of the Seventh SIAM Workshop on Combinatorial Scientific Computing pp 42–51 (*Preprint* `http://locus.siam.org/doi/pdf/10.1137/1.9781611974690.ch5`) URL `http://locus.siam.org/doi/abs/10.1137/1.9781611974690.ch5`